# Limitation of the Press-Schechter Formalism

Taihei YANO, Masahiro NAGASHIMA

and

Naoteru GOUDA

*Department of Earth and Space Science,*
*Faculty of Science, Osaka University,*
*Toyonaka, Osaka 560, Japan*

E-mail: yano@vega.ess.sci.osaka-u.ac.jp, masa@vega.ess.sci.osaka-u.ac.jp, gouda@vega.ess.sci.osaka-u.ac.jp

## ABSTRACT

The Press-Schechter(PS) formalism for the mass function of the collapsed objects are reanalyzed. The factor of two in the Press-Schechter formalism is convinced to be correct in the sharp $k$-space filter even when we use the another approach proposed by Jedamzik(1994) in the cloud-in-cloud problem, which is different from the previous approach by Peacock & Heavens(1990) and Bond et al.(1991). The spatial correlation of the density fluctuations, however, had been neglected in the cloud-in-cloud problem. The effects of this spatial correlation is analyzed by using the Jedamzik formalism and it is found that this effect alter the PS mass function especially on larger mass scales. Furthermore the exact formula of deriving mass function is shown. We also find that the probability of the overlap of the collapsed objects can be neglected on very small mass scales while it might not be neglected on other mass scales.
*Subject headings*: cosmology:theory-mass function-large scale structures-galaxy formations

# 1. INTRODUCTION

When we would like to understand the universe, the problem of the structure formation is one of the most important problems. In the standard cosmological model, we believe that the seeds of the structures are small density fluctuations in the early universe, and after the epoch of matter-radiation decoupling at redshift $z \sim 1000$, they have evolved and hierarchically clustered to the bound objects with extended magnitude of order of mass such as galaxies, clusters of galaxies and Lyman-$\alpha$ clouds, etc., via gravitational instability.

Therefore if the mass function for these bound objects would be known, we can in principle obtain the informations of the nature of the initial density fluctuations. It is, however, rather difficult to obtain completely analytic solutions for the problem of the structure formations since it is too complex to treat the non-linear phases of collapse to bound objects.

One of approaches to this problem, therefore, is N-body simulation. Since explicitly calculating gravitational force, it can follow the non-linear phases. However, by present computers, N-body simulations can work only in a limited dynamic range of mass.

Another analytic approach has been developed by Press and Schechter (1974; hereafter PS). They assumed that the initial density fields are expressed by the random Gaussian, and the bound objects have hierarchically formed from the initial density fluctuations . Then they proposed the formalism for counting the number density of objects with mass $M$ (i.e., the mass function). The PS mass function has been applied to various problems such as the number densities of galaxies and clusters of galaxies because of its *simple* form. However, the problem of their argument is that half of the mass, which is initially present in underdense regions, remains unaccounted for. With no clear reasons they *simply* multiplied the number density by a factor of two to take into account underdense regions since all of the mass must be considered. So it has been recognized that this factor of two is a crucial weak point of the PS analysis. To solve this problem, we must consider underdense regions which are embedded within overdense regions (it is called the "cloud-in-cloud" problem).

Peacock and Heavens (1990) and Bond et al. (1991) attempted to solve the cloud-in-cloud problem by using the peak formalism (Peacock and Heavens 1985; Bardeen et al. 1986). They considered the behavior of values of density peaks under varying the length of filters for smoothing the density field, that is, the probability that the underdense regions upcross to overdense regions in different length filters. In the sharp $k$-space filter, because of random phase, the behavior of the density field when summed up a wave with larger wavenumber $k$, is expressed by random walk. Only in this case, *surprisingly*, they obtained the same result as the PS formalism including the factor of two. In other filters instead of the sharp $k$-space filter, it is found that the factor 2 introduced in the PS formalism is not correct. As for as the cloud-in-cloud problem, these analyses are, however, insufficient : the spatial correlation functions of density fluctuations must be considered in order to solve correctly the cloud-in-cloud problem. In the previous works, however, they considered only the probability for the density fluctuations at one point in the space and neglected the spatial correlation of the density fluctuations in the cloud-in-cloud problem. In fact, it is found that the effect of the spatial correlation alter the PS formalism even in using the sharp $k$-space filter as we will show later.

Jedamzik (1994) approached the cloud-in-cloud problem with the formula using the integral equation of the mass function, which is different approach from the previous works. We believe it is useful approach to the cloud-in-cloud problem in taking into account the spatial correlation. But in his paper, he neglected the spatial correlation. Furthermore he used the wrong conditional probability about the condition of the isolated objects. So he misleadingly concluded that the factor of two in the PS formalism is incorrect even in the sharp $k$-space filter, which is inconsistent with Peacock & Heavens(1990) and Bond *et al.*(1991). As we will show later, however, we find that the factor of two can be derived in the sharp $k$-space filter even when we use the Jedamzik formalism in neglecting the spatial correlation, which is consistent with the previous works. Then we will conclude that the factor of two in the PS formalism is correct in the sharp $k$-space filter if the effects of the spatial correlation is neglected.

In this paper, we present how the effect of the spatial correlation in the cloud-in-cloud problem alter the PS formalism. And we show the new formula of mass function in taking into account the spatial correlation. In fact, it is very complicated to derive numerically the mass function with the new formula. And the form of the numerically derived mass function may not be simple enough for application to the various problems. However, we believe that it is very important task to analyze the limitation of the PS mass function. So we show approximately how the PS mass function deviate from the modified one in some special cases.

As for the derivation of mass function, we use the Jedamzik formalism with the correct conditional probability including the effect of the spatial correlation.

In §2, the PS formalism is briefly reviewed. And in §3, the Jedamzik formalism is shown. We present the modified formula with the spatial correlation in §4 and present how this effect of the spatial correlation alter the PS mass function. In §5 we also show the new formula of mass function using the constrained probability that the isolated object has maximum peak density, whose condition is necessary in order to correctly get the mass function in taking into account the spatial correlation. In §6, we devote to the conclusions and discussions. In the Appendix A, we show how the mass function derived by the new formula deviate from the PS mass function approximately and also suggest the troublesome effect of the overlap of the collapsed objects in the Appendix B.

## 2. THE PS FORMALISM

In this section, we briefly review the PS formalism. In the standard cosmological models, we assume that the initial small density fluctuations are the seeds of cosmic structures such as galaxies and they obey the random Gaussian distributions. Then the probability function for the Fourier components of density fluctuations, $\delta_{\mathbf{k}}$, is given by(Bardeen et al. 1986)

$$P(|\delta_{\mathbf{k}}|, \phi_{\mathbf{k}})d|\delta_{\mathbf{k}}|d\phi_{\mathbf{k}} = \frac{2|\delta_{\mathbf{k}}|}{P(k)}\exp\{-\frac{|\delta_{\mathbf{k}}|^2}{P(k)}\}d|\delta_{\mathbf{k}}|\frac{d\phi_{\mathbf{k}}}{2\pi}, \qquad (1)$$

where $\phi_{\mathbf{k}}$ is the random phase of $\delta_{\mathbf{k}}$ and $P(k)$ is the power spectrum.

In general, we assume that the density fluctuations associated with the objects of mass scale $M$ is given by the filtered density fluctuations defined as follows:

$$\delta_M(\mathbf{x}) = \frac{V}{(2\pi)^3} \int_0^\infty \delta_{\mathbf{k}} e^{i\mathbf{k}\cdot\mathbf{x}} W(k, R) d^3\mathbf{k},$$
$$R = (\frac{3M}{4\pi\bar{\rho}})^{1/3},$$

where $W(k, R)$ is an window function and $\bar{\rho}$ is the mean density.

It is called "sharp $k$-space filter" when we use the window function defined as

$$W(k, R) = \begin{cases} 1 & (k \leq k_c) \\ 0 & (k > k_c) \end{cases}$$
$$k_c = \frac{\pi}{R}$$

In this paper, we consider only this sharp $k$-space filter for simplicity. If we choose other filters, the results might change quantitatively.

The density fluctuation $\delta_M$ grows up as time goes and the objects with mass scale $M$ recollapse and vilialize when the linearly estimated $\delta_M$ grows to the critical density $\delta_c$. It is well known that the critical density $\delta_c \sim 1.69$ for the spherical collapse in the flat universe(Peebles 1993). Throughout this paper, we consider only the spherical symmetric case. Also in other cases, the critical density has the similar value. It is considered that the objects with mass scale $M$ vilialized at this epoch.

The probability $f(>\delta_c, M)$ to find a region of mass scale $M$ to be overdense by more than the critical amount $\delta_c$ is given by an integral over the tail of a Gaussian distribution function,

$$f(>\delta_c, M) = \int_{\delta_c}^{\infty} \frac{1}{\sqrt{2\pi\sigma^2(M)}} \exp\{-\frac{\delta^2}{2\sigma^2(M)}\} d\delta, \qquad (2)$$

because linearly estimated $\delta$ is the random Gaussian fields. Here the variance $\sigma^2(M)$ is estimated by sum up a variance of each Fourier components in the sharp $k$-space filter as follows:

$$\sigma^2(M) = \frac{V}{(2\pi)^3} \int_0^{k_c(M)} |\delta_k|^2 4\pi k^2 dk, \qquad (3)$$

$$k_c(M) = \pi/(\frac{3M}{4\pi\bar{\rho}})^{1/3}.$$

When we assume a simple power law, $P(k) \propto k^n$, the variance $\sigma^2(M)$ is given by

$$\sigma^2(M) = (\frac{M}{M_*})^{-\frac{(n+3)}{3}} \qquad (4)$$

where $M_*$ is the characteristic mass scale such that $\sigma^2(M_*) = 1$.

In the PS formalism, $f(>\delta_c, M)$ is considered to be proportional to the probability that a given point has ever been processed through a collapsed objects of mass scales $> M$. If $\delta_M > \delta_c$ for a given mass $M$ at a point, then it will have $\delta = \delta_c$ when filtered on some larger mass scale and so it will be counted as collapsed objects of the larger mass scale. So in this argument, it is assumed that the only objects which exist at a given epoch are those which have just collapsed($\delta = \delta_c$). Thus the mass function $n(M)$(defined such that $n(M)dM$ is the comoving number density of the collapsed objects in the range $dM$) is given by

$$\frac{M}{\bar{\rho}} n(M) = |\frac{df}{dM}| = \frac{\delta_c}{\sqrt{2\pi}\sigma} \frac{1}{M} |\frac{d\ln\sigma}{d\ln M}| \exp(-\frac{1}{2}\delta_c^2/\sigma^2).$$

In this argument, it only associates half of the mass of the initial density fluctuations with eventually to be collapsed. The problem is that half of the mass, which is initially present in underdense regions, remains unaccounted for.

Press and Schechter *simply* multiply the probability by a factor 2 with no clear physical reason. Thus,

$$n(M) = 2\frac{\bar{\rho}}{M}|\frac{df}{dM}| \equiv n_{ps}.$$

The factor of 2 has long been noted as weak point in the PS formalism. The problem of correctly counting the underdense regions which are embedded with overdense regions is called "cloud-in-cloud" problem. Peacock & Heavens(1990) and Bond et al.(1991) proposed a solution to the cloud-in-cloud problem in taking into account the probability, $P_{up}$, that subsequent filtering with larger scales might at some points result in having $\delta > \delta_c$ when at smaller filter $\delta < \delta_c$. *Surprisingly, in the sharp k-space filter, it is found that the factor of 2 in the PS formalism is correct.* However, the factor of 2 is incorrect when using the other filters.

## 3. JEDAMZIK FORMALISM

Jedamzik(1994) proposed the another approach to the cloud-in-cloud problem. We will explain this approach below: our goal is to compute the number density of *isolated* region. Isolated regions are regions which have collapsed but not included in yet larger regions overdense by $\delta_c$ or more. In particular, a region of mass scale $M$ will be only counted as an eventually vilialized object of mass $M$, if for any larger mass scale the average overdense of the region is below the critical density. Then the density fluctuation of the isolated regions is just the critical density $\delta_c$ because the regions with $\delta > \delta_c$ would be eventually counted as an object of the larger scale.

Now we define the conditional probability $P(M_1, M_2)$ of finding a region of mass scale $M_1$ overdense by $\delta_c$ or more, provided it is included in an isolated over dense region of mass scale $M_2 (> M_1)$. Then the volume ratio $f(> \delta_c, M_1)$, of finding a region of mass scale $M_1$ overdense by $\delta_c$ or more is given by the following integral form:

$$f(> \delta_c, M_1) = \int_0^\infty dM_2 n(M_2) \frac{M_2}{\bar{\rho}} P(M_1, M_2) \qquad (5)$$

Here $n(M)$ is the number density(i.e. mass function) of the isolated objects of mass scale $M$. If we can calculate $P(M_1, M_2)$ in any case, we can get a mass function $n(M)$ without worrying about the cloud-in-cloud problem in principle by using this formula(we call it the Jedamzik formalism).

By definition, the conditional probability must satisfy $P(M, M') = 0$ for $M > M'$. As if $P(M, M')$ would have the property such as $P(M, M') = 1$ for $M \leq M'$, then

$$|\frac{df}{dM}| = \frac{M}{\bar{\rho}} n(M),$$

where we use the relation such as $dP(M, M')/dM = -\delta(M - M')$. Then we get

$$n(M) = \frac{\bar{\rho}}{M} |\frac{df}{dM}|.$$

This is the original PS formula without the factor of 2. In general, however, this case could not appear in real.

In the sharp $k$-space filter, we find

$$P(M, M') = \int_{\delta_c}^{\infty} d\delta_M \frac{1}{\sqrt{2\pi}} \frac{1}{\sigma_{sub}} \exp\{-\frac{1}{2} \frac{(\delta_M - \delta_c)^2}{\sigma_{sub}^2}\}$$
$$= \frac{1}{2},$$

where

$$\sigma_{sub}^2 = (\frac{M}{M_*})^{-(1+\frac{n}{3})} - (\frac{M'}{M_*})^{-(1+\frac{n}{3})}.$$

Here we use the fact that the region with mass $M'$ is the isolated one, then $\delta_{M'} = \delta_c$.

Then $dP(M, M')/dM = -\frac{1}{2}\delta(M - M')$. Hence, from eq.(5),

$$|\frac{df}{dM}| = \frac{1}{2} \frac{M}{\bar{\rho}} n(M)$$

and so

$$n(M) = 2 \frac{\bar{\rho}}{M} |\frac{df}{dM}|.$$

This is just the PS mass function including the factor of 2. Then we proved that in the sharp $k$-space filter, the factor of 2 is correct even when we use the Jedamzik formalism, which is different approach from the previous works (in Jedamzik's paper(1994), he used the wrong conditional probability in the sharp $k$-space filter. So he derived the wrong result that the factor of 2 is incorrect even using the sharp $k$-space filter). This Jedamzik formula is very useful to analyze the mass function with the effect of the spatial correlation functions in the cloud-in-cloud problem. If we can get the conditional probability $P(M, M')$ in taking into account the spatial correlation, we can in principle estimate the mass function by solving the integral equation (eq.(5)).

## 4. SPATIAL CORRELATION

In the previous works, they considered the probability for the density fluctuations only at one point. In reality, however, the object of mass scale $M$ will have a finite size and will include the other objects in this finite region. So we must consider the two-point correlation of the density fluctuations in order to analyze whether the objects of smaller mass scale $M_1$ are included in the finite region of the object with a larger mass scale $M_2$. Then we believe that it is necessary to consider the spatial correlation of the density fluctuations in order to solve the cloud-in-cloud problem correctly.

Now we consider the following conditional probability for simplicity in order to analyze only the effects of the spatial correlation on the PS formalism at first(in the next section, we will show the exact conditional probability for estimating the *correct* mass function while the form of the probability is complicated). In estimating the conditional probability $P(M_1, M_2)$, at first, we consider the conditional probability, $P(r, M_1, M_2)$, of finding a region of mass scale $M_1$ overdense by $\delta_c$ or more at distance $r$ from the center ($r = 0$) of the isolated object of mass scale $M_2$, provided the object with mass $M_1$ is included in a object of mass scale $M_2 (> M_1)$ with a finite size and $\delta_{M_2} = \delta_c$ at $r = 0$. As we will show later, it is enough good to consider this conditional probability in order to analyze the effects of the spatial correlation. while in this probability the condition that the $\delta_{M_2}$ has the maximum peak density is neglected(see §5).

We find that $P(r, M_1, M_2)$ is equivalent to the constrained probability defined as follows(Bardeen et al. 1986):

$$P(r, M_1, M_2) = P(\delta_{M_1}(r) > \delta_c | \delta_{M_2} = \delta_c)$$
$$= \frac{1}{\sqrt{2\pi(1-\varepsilon^2(r))}} \int_{\nu_{1c}}^{\infty} \exp[\frac{-(\nu_1 - \varepsilon(r)\nu_{2c})^2}{2(1-\varepsilon^2(r))}] d\nu_1 \qquad (6)$$

where $\nu_1, \nu_2$ have a unity variance define as

$$\nu_1 \equiv \frac{\delta_1}{\sigma_1}, \nu_2 \equiv \frac{\delta_2}{\sigma_2}, \sigma_1 \equiv \sigma(M_1), \sigma_2 \equiv \sigma(M_2), \nu_{1c} = \frac{\delta_c}{\sigma_1}, \nu_{2c} = \frac{\delta_c}{\sigma_2} \qquad (7)$$

and $\varepsilon(r)$ is defined by

$$\varepsilon(r) \equiv \varepsilon_0 \alpha(r) \equiv \frac{\sigma_h^2}{\sigma_1 \sigma_2} \alpha(r), \qquad (8)$$

which corresponds to the two-point correlation function. Here $\sigma_h$ is a cross-correlation of the density fluctuations with the mass scale $M_1$ and the mass scale $M_2$ at the same point. When we use a sharp $k$-space filter, the cross-correlation is the same as the variance of the density fluctuation filtered with the larger mass scale. Therefore, $\sigma_h$ is equal to the variance $\sigma_2$. Then, in this case, $\varepsilon_0 = \sigma_2/\sigma_1$.

The conditional probability is rewritten by

$$P(r, M_1, M_2) = \frac{1}{\sqrt{2\pi}} \int_{\beta(r)}^{\infty} e^{-\frac{y^2}{2}} dy, \qquad (9)$$

where

$$\beta(r) \equiv \frac{\nu_{1c} - \varepsilon(r)\nu_{2c}}{\sqrt{1-\varepsilon^2(r)}} = \frac{\delta_c}{\sigma_1} \frac{1-\alpha(r)}{\sqrt{1-\varepsilon_0^2 \alpha^2(r)}},$$

$$y \equiv \frac{\nu_1 - \varepsilon_0 \alpha(r)\nu_{2c}}{\sqrt{1-\varepsilon_0^2 \alpha^2(r)}}.$$

Now we consider the spatial averaged conditional probability $P(M_1, M_2)$ de-

fined as

$$P(M_1, M_2) \equiv \int_0^R P(r, M_1, M_2) 4\pi r^2 dr / \int_0^R 4\pi r^2 dr$$
$$\equiv \frac{1}{\sqrt{2\pi}} \int_{\bar{\beta}}^{\infty} e^{-\frac{y^2}{2}} dy, \qquad (10)$$

where $R \sim R_2 = (3M_2/4\pi\bar{\rho})^{1/3}$ and

$$\bar{\beta} \equiv \frac{1}{\sqrt{1-\varepsilon_0^2 \bar{\alpha}^2}} \frac{\delta_c}{\sigma_1}(1-\bar{\alpha}) = \frac{1}{\sqrt{1-(\frac{M_2}{M_1})^{-\frac{n+3}{3}}\bar{\alpha}^2}} (\frac{\delta_c}{\sigma_1})(1-\bar{\alpha}).$$

The above equations give the definitions of the constant values, $\bar{\beta}$ and $\bar{\alpha}$. Then after when we calculate this conditional probability in any case, we can solve the mass function by substituting this probability into the integral equation (5). If we neglect the spatial correlation, then $\bar{\alpha} = \alpha(0) = 1$ and so $P(M_1, M_2) = 1/2$. Hence again we can get the PS mass function.

In practice, it may be a little complicated to solve numerically the mass function with the integral equation. However we can get approximately the mass function with good accuracy as shown below: we find that $\bar{\alpha}$ is between 0 and 1 and its value depends on the power spectrum, but $\bar{\alpha}$ is approximately around 0.5 (for $-2 \lesssim n \lesssim 2$). We consider these cases for that $\bar{\alpha}^2$ is much less than unity. In these cases, we can approximately get

$$\bar{\beta} \sim \frac{\delta_c}{\sigma_1}(1-\bar{\alpha}). \qquad (11)$$

The error at $M_1 \sim M_2$ is at most around 10%. Then we get the following conditional probability with good accuracy:

$$P(M, M') \begin{cases} \equiv g(M) = \frac{1}{\sqrt{2\pi}} \int_{\bar{\beta}}^{\infty} e^{-\frac{y^2}{2}} dy & (M \leq M') \\ = 0 & (M > M') \end{cases} \qquad (12)$$

where $\bar{\beta} = \frac{\delta_c}{\sigma(M)}(1-\bar{\alpha})$.

Using the Jedamzik formalism(eq.(5)),

$$\frac{f(>\delta_c, M)}{g(M)} = \frac{1}{\bar{\rho}} \int_0^\infty dM' M' n(M') \frac{P(M, M')}{g(M)}. \quad (13)$$

Here,

$$\frac{P(M, M')}{g(M)} = \begin{cases} 1 & (M \leq M') \\ 0 & (M > M') \end{cases}$$

Then we can get the following equation,

$$|\frac{d}{dM}\{\frac{f}{g}\}| = \frac{1}{\bar{\rho}} M n(M) \quad (14),$$

because $d\{\frac{P}{g}\}/dM = -\delta(M - M')$.

Now we estimate the mass function in two extreme cases. First of them, we consider the small mass scale ranges of mass function $(M \ll M_*)$. In this case, we can use the relation such as $\delta_c/\sigma(M) \ll 1$. Then

$$f = \frac{1}{\sqrt{2\pi}} \int_{\frac{\delta_c}{\sigma(M)}}^\infty e^{-\frac{y^2}{2}} dy \sim \frac{1}{2} - \frac{1}{\sqrt{2\pi}} \frac{\delta_c}{\sigma(M)}$$

$$g = \frac{1}{\sqrt{2\pi}} \int_{\frac{\delta_c}{\sigma(M)}(1-\bar{\alpha})}^\infty e^{-\frac{y^2}{2}} dy \sim \frac{1}{2} - \frac{1}{\sqrt{2\pi}} \frac{\delta_c}{\sigma(M)}(1-\bar{\alpha})$$

In order to see the difference between the modified mass function and the PS mass function $n_{ps}$, we see the ratio of $d(f/g)/dM$ with $2df/dM$.

$$\frac{d}{dM}\{\frac{f}{g}\}/\frac{2df}{dM}|_{f=1/2} = \frac{d(\frac{f}{g})}{2df}|_{f=1/2}$$

$$= \frac{\bar{\alpha}/2}{2\{(1-\bar{\alpha})f + \bar{\alpha}/2\}^2}|_{f=1/2} = \bar{\alpha} \quad (15)$$

Next we consider the larger mass scale ranges of the mass function $(M \gg M_*)$.

In this case, $\delta_c/\sigma(M) \gg 1$. We can use the next approximate expression,

$$\frac{1}{\sqrt{2\pi}}\int_{\bar{\beta}}^{\infty} e^{\frac{-x^2}{2}}dx \sim \frac{1}{\sqrt{2\pi}}e^{-\frac{\bar{\beta}^2}{2}}\frac{1}{\bar{\beta}} \quad (\bar{\beta} \gg 1)$$

So, the ratio is

$$\frac{d}{dM}\{\frac{f}{g}\}/\frac{2df}{dM} = \frac{d}{dx}\{\frac{f}{g}\}/\frac{2df}{dx}$$
$$= \frac{1}{2}\sqrt{2\pi}(1-\bar{\alpha})\{1-(1-\bar{\alpha})^2\}x\exp\{\frac{1}{2}x^2(1-\bar{\alpha})^2\}, \quad (16)$$

where $x \equiv \delta_c/\sigma = (M/M_*)^{\frac{1}{2}+\frac{n}{6}}\delta_c$

From the above equations (15) and (16), the modified mass function is given as follows:

$$n(M) = \frac{\bar{\rho}}{M}|\frac{d}{dM}\{\frac{f}{g}\}|$$

$$\sim \begin{cases} \bar{\alpha}n_{ps} & (M \ll M_*) \\ \frac{\sqrt{2\pi}}{2}\bar{\alpha}(1-\bar{\alpha})(2-\bar{\alpha})(\frac{M}{M_*})^{\frac{1}{2}+\frac{n}{6}}\delta_c\exp\{\frac{1}{2}(\frac{M}{M_*})^{(1+\frac{n}{3})}\delta_c^2(1-\bar{\alpha})^2\}n_{ps} & (M \gg M_*) \end{cases} \quad (17)$$

As we can see, the deviation from PS mass function $n_{ps}$ is within factor $\bar{\alpha}$ on smaller mass scales($M \ll M_*$). However on larger mass scale ranges of mass function($M \gg M_*$), we find large deviation from the PS mass function in some cases. Of course, it must be noted that the characteristic mass $M_*$ change as time goes while $M_*$ is initially small.

## 5. EXACT FORMULA OF MASS FUNCTION

In the previous section, we showed how the effect of the spatial correlation alter the PS mass function. Strictly speaking, the conditional probability $P(M_1, M_2)$ shown in §4 is not sufficient to get the *exact* mass function. The *isolated* object with a large mass $M_2$ must have the *maximum peak density* with $\delta_{M_2} = \delta_c$ at $r = 0$.

So we need to consider the conditional probability of finding a region of mass scale $M_1$ overdense by $\delta_c$ or more at distance $r$, provided the object with mass scale is included in an *isolated* object of mass scale $M_2$ *with the maximum peak density* of $\delta_{M_2} = \delta_c$ at $r = 0$. Then we must use the following constrained probability,

$$P(r, M_1, M_2 | \text{peak}) = P(\delta_{M_1}(r) > \delta_c | \delta_{M_2} = \delta_c, \text{peak})$$

$$= \frac{1}{\sqrt{2\pi Q_2}} \int_{\nu_{1c}}^{\infty} \exp(\frac{-1}{2} \frac{Q_1}{Q_2}) d\nu_1,$$

$$Q_1 = [\nu_1 - \varepsilon_{0r}\{\frac{1 - \frac{9\varepsilon_{2r}\varepsilon_{02}}{5\varepsilon_{0r}}}{1 - \frac{9}{5}\varepsilon_{02}^2}\}\nu_2 + \frac{3}{5}\frac{\varepsilon_{0r}\varepsilon_{02} - \varepsilon_{2r}}{1 - \frac{9}{5}\varepsilon_{02}^2}(\nu_6 + \nu_7 + \nu_8)]^2,$$

$$Q_2 = 1 - \varepsilon_{0r}^2 \frac{1 + \frac{9}{5}\frac{\varepsilon_{2r}^2}{\varepsilon_{0r}^2} - \frac{18\varepsilon_{2r}\varepsilon_{02}}{5\varepsilon_{0r}}}{1 - \frac{9\varepsilon_{02}^2}{5}},$$

where $\nu_i$ and $\nu_{1c}$ are defined as normalized variables:

$$\nu_i = \frac{V_i}{\langle V_i^2 \rangle^{\frac{1}{2}}},$$

$$\nu_{1c} = \frac{\delta_c}{\sigma_r},$$

$$\mathbf{V} = (\delta(\mathbf{r}), \delta, \delta_i', \delta_{ii}'', \delta_{ij}'').$$

And $\varepsilon$ is correlation coefficient,

$$\varepsilon_{0r} = \frac{\sigma_{0r}^2}{\sigma_0 \sigma_r}$$

$$\varepsilon_{02} = -\frac{\sqrt{5}}{3} \frac{\sigma_1^2}{\sigma_0 \sigma_2}$$

$$\varepsilon_{2r} = -\frac{\sqrt{5}}{3} \frac{\sigma_{1r}^2}{\sigma_2 \sigma_r}$$

where $\sigma$ means variance,

$$\sigma_m^2 = \frac{4\pi V}{(2\pi)^3} \int_0^{\infty} W^2(k, R_2) |\delta_k|^2 k^{(2m+2)} dk$$

$$\sigma_{mr}^2 = \frac{4\pi V}{(2\pi)^3} \int_0^{\infty} W(k, R_1) W(k, R_2) |\delta_k|^2 k^{(2m+2)} \frac{\sin(kr)}{kr} dk$$

$$\sigma_r^2 = \frac{4\pi V}{(2\pi)^3} \int_0^{\infty} W^2(k, R_1) |\delta_k|^2 k^2 dk.$$

The derivation of the above probability is explained in Appendix A. Thus we get the conditional probability $P(M_1, M_2|\text{peak})$ after averaging spatially the above probability. This is the *exact* procedure of getting the mass function. In practice, however, it is much complicated to solve numerically the mass function. But we can see how the mass function with this formalism deviate from the PS mass function approximately. The detailed method is explained in the Appendix A. The result is that the deviation is similar to that shown in §4(eq.17), in which we neglect the condition that the object with mass $M_2$ has *maximum peak* density. So we conclude that the deviation is within factor on smaller mass scales, however, it is very large on larger mass scales in some cases(see eq.(A-15)).

Furthermore we investigate the probability that the collapsed objects would overlap. If the overlap would occur, we could not account the number of the objects because we can not determine whether the overlapped objects merge to one larger mass object or fragment into two small objects in our statistical argument. In this case it is necessary to analyze dynamically the effect of the overlap. As shown in the Appendix B, by using the exact formula of mass function, we find that the probability of the overlap can be neglected on very small mass scales while it might not be neglected on other mass scales. The effect of the overlap especially on medium mass scales gives a serious problem to estimate the mass function in the statistical argument.

## 6. CONCLUSIONS AND DISCUSSIONS

We have reanalyzed the PS formalism. When we neglect the spatial correlation of the density fluctuations in the cloud-in-cloud problem, we can get just the PS mass function including the factor of two in the sharp $k$-space filter by using the Jedamzik formalism, which is different approach from the previous one used by Peacock & Heavens(1990) and Bond et al.(1991). However we believe that it is very important to take into account the effect of the spatial correlation in order to estimate correctly the mass function in real. The Jedamzik formalism is very useful to estimate the mass function when we consider the effect of the spatial correlation in the cloud-in-cloud problem. By using this formalism, we showed how the effect of the spatial correlation alter the PS mass function. Furthermore, we showed the exact formula of finding the mass function in taking into account the exact probability that the isolated object has the maximum peak density. Although it is difficult to solve numerically mass function with this formula in practice, we can approximately show that the deviation of the modified mass function from the PS mass function is similar in the case that we consider only the effect of the spatial correlation. We conclude that while the deviation on smaller mass scales is within factor $\bar{\alpha}$, whose value depends on the power spectrum, it is very large on larger mass scales in some cases. So we warn us that we must be careful about the limitation of the PS formalism when we apply the PS mass function to some problems. Furthermore we found that we can neglect the probability of overlap between the collapsed objects on very small mass scales while it might not be neglected on other mass scales. This problem of the overlap especially on medium mass scales is a serious one to estimate the mass function in the statistical argument such as the PS formalism or our formalism.

## ACKNOWLEDGMENT

We are grateful to K.Okoshi and H.Susa for stimulating discussions. This work was supported in part by the Grants-in-Aid 06640352 for the Scientific Research Fund from the Ministry of Education, Science and Culture of Japan.

## APPENDIX A

In §5, we mentioned the exact formula of mass function. Here, we express the conditional probability on which the exact formula is based. And we will also show how the mass function derived by the exact formula deviate from the PS mass function *approximately*.

### (a) Multivariate Gaussian distribution function

When m-variables are associated with a Gaussian field, the probabilities given by

$$P(\mathbf{V})d\mathbf{V} = \frac{e^{(-\frac{Q}{2})}}{\sqrt{(2\pi)^m \det(\mathbf{M})}} d\mathbf{V}, \qquad (A-1)$$

where

$$Q = \mathbf{V}^T \mathbf{M}^{-1} \mathbf{V} \qquad (A-2)$$

$$\mathbf{V} = (y_1, \ldots, y_m) \qquad (A-3)$$

$$M_{ij} = \langle (y_i - \langle y_i \rangle)(y_j - \langle y_j \rangle) \rangle \qquad (A-4)$$

We use the brackets $\langle \rangle$ as the ensemble mean of the universe, but in practice, assuming homogeneity and ergodicity in space, we take it as the spatial mean, so $\langle y_i \rangle$ corresponds to the spatial mean of $y_i$ and $M_{ij}$ is covariance between $y_i$ and $y_j$. Now, because we want the conditional probability of maximum peak, we must calculate the first and second derivative of the density perturbation $\delta$. So we need to consider the $11 \times 11$ covariance matrix $\mathbf{M}$, where $\mathbf{V} = (\delta(\mathbf{r}), \delta, \delta'_i, \delta''_{ii}, \delta''_{ij}), i, j = 1, 2, 3, i \neq j$, and $'$ means derivative with respect to $\mathbf{r}$. The non-zero components of this matrix are as follows (of course, $\mathbf{M}$ is symmetric matrix):

$$M_{11} = \sigma_r^2$$

$$M_{22} = \sigma_0^2$$

$$M_{33} = M_{44} = M_{55} = \frac{\sigma_1^2}{3}$$

$$M_{66} = M_{77} = M_{88} = \frac{\sigma_2^2}{5}$$

$$M_{99} = M_{10,10} = M_{11,11} = M_{67} = M_{78} = M_{86} = \frac{\sigma_2^2}{15}$$

$$M_{26} = M_{27} = M_{28} = -\frac{\sigma_1^2}{3}$$

$$M_{21} = \sigma_{0r}^2$$

$$M_{16} = M_{17} = M_{18} = -\frac{\sigma_{1r}^2}{3}, \qquad (A-5)$$

where

$$\sigma_m^2 = \frac{4\pi V}{(2\pi)^3} \int_0^\infty W^2(k, R_2) |\delta_k|^2 k^{(2m+2)} dk$$

$$\sigma_{mr}^2 = \frac{4\pi V}{(2\pi)^3} \int_0^\infty W(k, R_1) W(k, R_2) |\delta_k|^2 k^{(2m+2)} \frac{\sin(kr)}{kr} dk$$

$$\sigma_r^2 = \frac{4\pi V}{(2\pi)^3} \int_0^\infty W^2(k, R_1) |\delta_k|^2 k^2 dk, \qquad (A-6)$$

and $W(k, R)$ is a window function and we find that $\sigma_{0r}^2 = \sigma_h^2 \alpha(r)$ which is defined in §4. Throughout this paper, we use the sharp-$k$ filter function, therefore, $W(k, R)$ is given by

$$W(k, R) = \begin{cases} 1 & (k \leq k_c(R)) \\ 0 & (k > k_c(R)) \end{cases}$$

By calculating these components, we obtain the multivariate Gaussian distribution function $P(\mathbf{V})$.

(b) Conditional probability of maximum peak

Now, the probability which we want is the conditional probability $P(r, M_1, M_2 | peak)$ (see §5). From Bayes' theorem, this probability is related to multivariate Gaussians as follows:

$$P(r, M_1, M_2 | peak) = \frac{1}{P(\mathbf{V_0})} \int_{\nu_{1c}}^{\infty} P(\mathbf{V}) d\nu_1, \qquad (A-7)$$

where

$$\mathbf{V} = (\delta(\mathbf{r}), \delta, \delta'_i, \delta''_{ii}, \delta''_{ij})$$

$$\mathbf{V_0} = (\delta, \delta'_i, \delta''_{ii}, \delta''_{ij})$$

From eqs.(A-1) and (A-7), we obtain

$$P(r, M_1, M_2 | peak) = P(\nu_1 > \nu_{1c} | \nu_2, \ldots, \nu_{11})$$

$$= \frac{1}{\sqrt{2\pi Q_2}} \int_{\nu_{1c}}^{\infty} \exp\left(\frac{-1}{2} \frac{Q_1}{Q_2}\right) d\nu_1, \quad (A-8)$$

$$Q_1 = [\nu_1 - \varepsilon_{0r} \{\frac{1 - \frac{9\varepsilon_{2r}\varepsilon_{02}}{5\varepsilon_{0r}}}{1 - \frac{9}{5}\varepsilon_{02}^2}\}\nu_2 + \frac{3}{5}\frac{\varepsilon_{0r}\varepsilon_{02} - \varepsilon_{2r}}{1 - \frac{9}{5}\varepsilon_{02}^2}(\nu_6 + \nu_7 + \nu_8)]^2,$$

$$Q_2 = 1 - \varepsilon_{0r}^2 \frac{1 + \frac{9}{5}\frac{\varepsilon_{2r}^2}{\varepsilon_{0r}^2} - \frac{18\varepsilon_{2r}\varepsilon_{02}}{5\varepsilon_{0r}}}{1 - \frac{9\varepsilon_{02}^2}{5}},$$

where $\nu_i$ and $\nu_r^c$ are defined as normalized variable,

$$\nu_i \equiv \frac{V_i}{\langle V_i^2 \rangle^{\frac{1}{2}}},$$

$$\nu_{1c} \equiv \frac{\delta_c}{\sigma_r}. \qquad (A-9)$$

And $\varepsilon$ is correlation coefficient,

$$\varepsilon_{0r} = \varepsilon(r) = \frac{\sigma_{0r}^2}{\sigma_0 \sigma_r}$$

$$\varepsilon_{02} = -\frac{\sqrt{5}}{3} \frac{\sigma_1^2}{\sigma_0 \sigma_2}$$

$$\varepsilon_{2r} = -\frac{\sqrt{5}}{3} \frac{\sigma_{1r}^2}{\sigma_2 \sigma_r} \qquad (A-10)$$

From the equations (A-6)~(A-10), we can rewrite the conditional probability as follows:

$$P(r, M_1, M_2 | peak) =$$
$$\frac{1}{\sqrt{2\pi(1 - \varepsilon_{0r}^2(1 + \Delta_3(r)))}} \times$$
$$\int_{\nu_{1c}}^{\infty} \exp\{-\frac{1}{2} \frac{(\nu_1 - (1 + \Delta_1(r) - 3c\Delta_2)\varepsilon_{0r}\nu_{2c})^2}{1 - \varepsilon_{0r}^2(1 + \Delta_3(r))}\} d\nu_1 \quad (A-11)$$

where

$$1 + \Delta_1(r) \equiv 1 + \frac{1}{4}(n+3)(n+7)(1 - \frac{\alpha_{n+2}}{\alpha_n})$$

$$\Delta_2 \equiv \sqrt{\frac{1}{80}(n+3)(n+7)(n+5)^2(1 - \frac{\alpha_{n+2}}{\alpha_n})}$$

$$1 + \Delta_3(r) \equiv 1 + \frac{1}{4}(n+3)(n+7)(1 - \frac{\alpha_{n+2}}{\alpha_n})^2$$

$$\nu_6 + \nu_7 + \nu_8 \equiv -3c\nu_c$$

Here $\alpha_n$ is defined as follows:

$$\alpha_n \equiv \frac{\sigma_h^2(r)}{\sigma_h^2(0)} = \int |\delta_k|^2 4\pi k^2 \frac{\sin kr}{kr} |W_1||W_2| dk / \int |\delta_k|^2 4\pi k^2 |W_1||W_2| dk.$$

where $W_1 \equiv W(k, R_1)$ and $W_2 \equiv W(k, R_2)$. When the isolated object has the maximum peak density, $\nu_6, \nu_7$ and $\nu_8$ must be negative value and so $c$ must be positive. And we also find that its value is around 1 for any $n$.

Furthermore, the conditional probability is rewritten by

$$P(r, M_1, M_2 | peak) = \frac{1}{\sqrt{2\pi}} \int_{\gamma(r)}^{\infty} e^{-\frac{y^2}{2}} dy, \quad (A-12)$$

where

$$\gamma(r) \equiv \frac{\nu_{1c} - \varepsilon_{0r}(1 + \Delta_1(r) - 3c\Delta_2)\nu_{2c}}{\sqrt{1 - \varepsilon_{0r}^2(1 + \Delta_3(r))}} = \frac{\delta_c}{\sigma_1} \frac{1 - \alpha_n(r)(1 + \Delta_1(r) - 3c\Delta_2)}{\sqrt{1 - \varepsilon_0^2 \alpha_n^2(r)(1 + \Delta_3(r))}}$$

$$= \frac{\delta_c}{\sigma_1} \frac{1 - s(r)}{\sqrt{1 - \varepsilon_0^2 \alpha_n^2(r)(1 + \Delta_3(r))}}$$

$$s(r) \equiv (1 + \Delta_1(r) - 3c\Delta_2(r))\alpha_n(r)$$

As in §4, we get the spatial averaged probability $P(M_1, M_2|peak)$ with the constant value defined as

$$\bar{\gamma} \equiv \frac{1}{\sqrt{1 - (\frac{M_2}{M_1})^{-\frac{n+3}{3}}\bar{\alpha}^2(1+\bar{\Delta}_3)}}(\frac{\delta_c}{\sigma_1})(1 - \bar{s}).$$
$$\bar{s} \equiv (1 + \bar{\Delta}_1 - 3c\bar{\Delta}_2)\bar{\alpha} \qquad (A - 13)$$

Also in this case as in §4, we can approximate $\bar{\gamma}$ as follows with good accuracy(within 10% error around $M_1 \sim M_2$) because $\bar{\alpha} \sim 0.5$ and also we find that $\bar{\Delta}_3 \ll 1$:

$$\bar{\gamma} \sim \frac{\delta_c}{\sigma_1}(1 - \bar{s}). \qquad (A - 14)$$

We find that $\bar{s}$ is a little smaller than $\bar{\alpha}$.

Hereafter, we can use the same procedure as in §4, in which the difference between the exact mass function derived by this formula and the PS mass function are deduced. Then we conclude that the deviation from the PS mass function is similar in the results shown in §4 while the factor $\bar{s}$ is a little different from $\bar{\alpha}$:

$$n(M) = \frac{\bar{\rho}}{M}|\frac{d}{dM}\{\frac{f}{g}\}|$$

$$\sim \begin{cases} \bar{s}n_{ps} & (M \ll M_*) \\ \frac{\sqrt{2\pi}}{2}\bar{s}(1-\bar{s})(2-\bar{s})(\frac{M}{M_*})^{\frac{1}{2}+\frac{n}{6}}\delta_c\exp\{\frac{1}{2}(\frac{M}{M_*})^{(1+\frac{n}{3})}\delta_c^2(1-\bar{s})^2\}n_{ps} & (M \gg M_*) \end{cases} \qquad (A - 15)$$

## APPENDIX B

In this appendix, we consider the effect of the overlap between the collapsed objects.

We assume that there is an isolated collapsed object of mass scale $M_2$ overdense by just $\delta_c$ with a maximum peak density at $r = 0$. If an isolated object of mass scale $M_1 (\leq M_2)$ locates at $R_2 - R_1 \lesssim r \lesssim R_2 + R_1$, the overlap between the objects of mass $M_2$ and $M_1$ occur.

Now we consider the condition that the object of mass scale $M_1$, whose scale is between mass scale 0 and $M_2$, has the density of $\delta_c$ at $R_2 - R_1 \lesssim r \lesssim R_2 + R_1$, that is, the object of mass scale $M_1$ overlap with the object of mass scale $M_2$. So we consider the following conditional probability of finding a region of mass scale between 0 and $M_2$ overdense by just $\delta_c$ at the distance $R_2 - R_1 \lesssim r \lesssim R_2 + R_1$ from the center of the isolated object of mass scale $M_2$.

$$P(\delta = \delta_c, 0 < M < M_2) = \{P(R_2 - R_1 \lesssim r \lesssim R_2 + R_1, 0, M_2 | peak) -$$
$$P(R_2 - R_1 \lesssim r \lesssim R_2 + R_1, M_2, M_2 | peak)\}$$
$$/\{1 - P(R_2 - R_1 \lesssim r \lesssim R_2 + R_1, M_2, M_2 | peak)\},$$

where

$$P(R_2 - R_1 \lesssim r \lesssim R_2 + R_1, M_1, M_2 | peak) = \int_{\gamma(r)}^{\infty} e^{-\frac{y^2}{2}} dy$$

$$\gamma(r) = \frac{\delta_c}{\sigma(M_1)} (1 - s(r))$$

when the mass scale $M_2$ is much smaller than the characteristic scale $M_*$, both conditional probability $P(R_2 - R_1 \lesssim r \lesssim R_2 + R_1, 0, M_2 | peak)$ and $P(R_2 - R_1 \lesssim r \lesssim R_2 + R_1, M_2, M_2 | peak)$ can be nearly equal to $\frac{1}{2}$ because both $\sigma(M_1)$ and $\sigma(M_2)$ are much larger than unity and then $\gamma(r)$ is nearly equals to 0 in each case. So the conditional probability $P(\delta = \delta_c, 0 < M < M_2)$ can be neglect for very small mass scales. But if the mass scale $M_2$ is not much smaller than $M_*$, the probability of the overlap might not be neglected.

However, the objects with much smaller mass scale $M_1 (\ll M_2)$ dominantly contribute to the overlap with the object of mass scale $M_2$ as we can see from the above probability. So the mass of the object with mass scale $M_2 (\gg M_*)$ is not changed so much by the overlap with the objects of such very small mass scales. Hence, as for the very large mass scale $M_2$, the mass function of $M_2$ itself might not be much affected by the effect of the overlap. The effect of the overlap on medium mass scales might give a serious problem for estimating the mass function.